\documentclass[journal=aesccq,manuscript=article]{achemso}

\usepackage[version=3]{mhchem} 
\usepackage{xcolor}

\author{Vincent J. Esposito}
\email{vincent.j.esposito@nasa.gov}
\affiliation[NASA Ames Research Center]
{Astrophysics Branch, NASA Ames Research Center, MS 245-6, Moffett Field, CA 94035, USA}

\author{Ryan C. Fortenberry}
\affiliation{Department of Chemistry \& Biochemistry, University of Mississippi, University, MS 38677-1848, USA}

\author{Christiaan Boersma}
\affiliation{Astrophysics Branch, NASA Ames Research Center, MS 245-6, Moffett Field, CA 94035, USA}

\author{Louis J. Allamandola}
\affiliation{Astrophysics Branch, NASA Ames Research Center, MS 245-6, Moffett Field, CA 94035, USA}

\title[Infrared Absorption Spectrum of CN-PAHs]
  {The High-Resolution Far- to Near-Infrared Anharmonic Absorption Spectra of Cyano-Substituted Polycyclic Aromatic Hydrocarbons from 300 - 6200~cm$^{-1}$}


\begin{document}

\begin{tocentry}

 \includegraphics{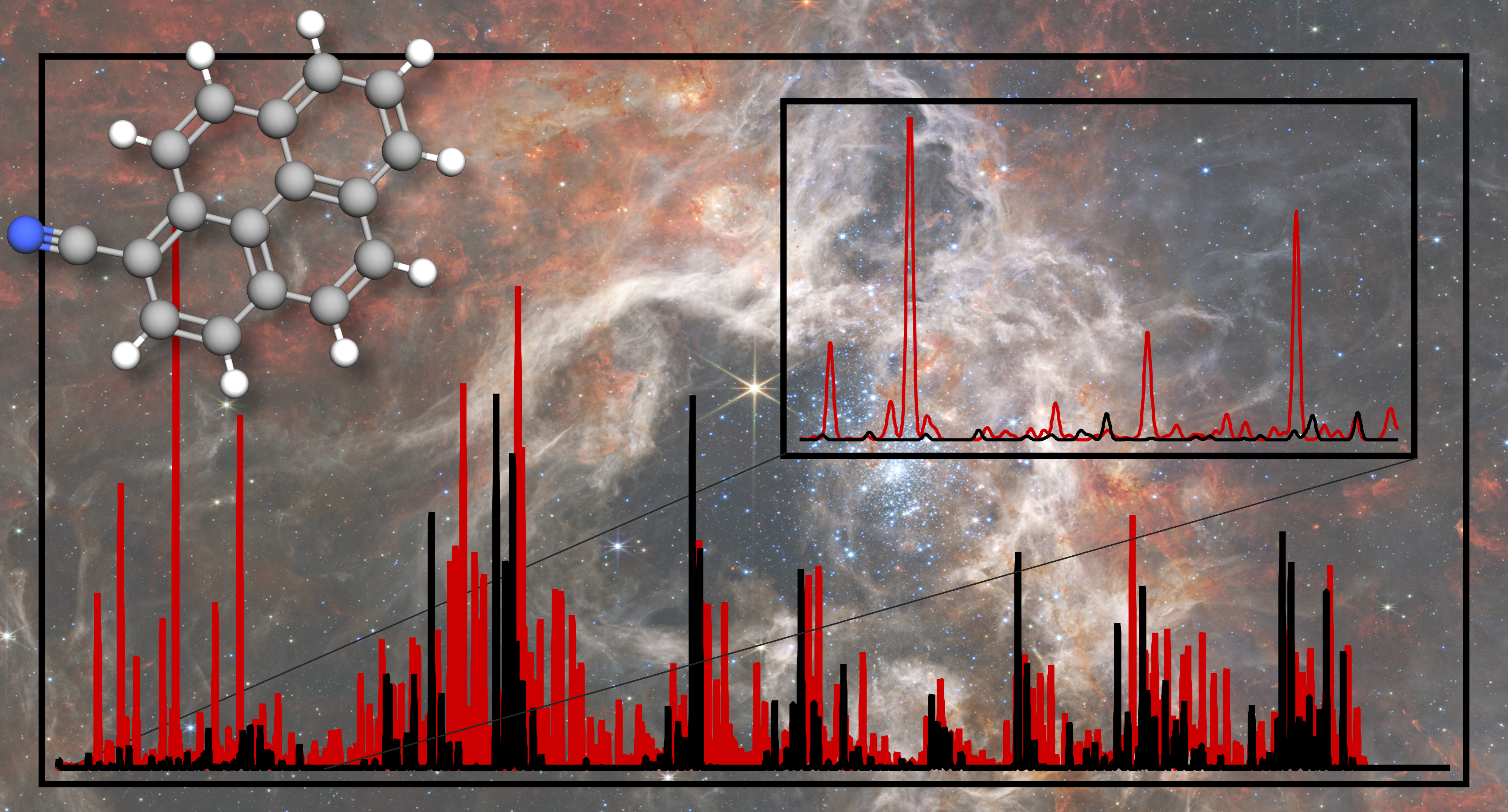}
 
\end{tocentry}

\begin{abstract}

Cyano-substituted polycyclic aromatic hydrocarbons (CN-PAHs) may contribute to the emission detected in the 7 - 9 \textmu m (1430 - 1100~cm$^{-1}$) and 11 - 15 \textmu m (900 - 670~cm$^{-1}$) regions of astronomical IR spectra. Anharmonic quantum chemical computations of 17 CN-PAH isomers for 4 small PAHs and Benzene reveal strong, broad absorption features across the entire 300 - 6200~cm$^{-1}$ (33 - 1.6 \textmu m) frequency range. In particular, when a FWHM of 15~cm$^{-1}$ is applied, the composite CN-PAH spectrum is almost indistinguishable from the unsubstituted-PAH spectrum. At high resolution, however, the infrared absorption spectra reveal unique, identifiable features of CN-PAHs in the 700 - 950, 1100 - 1300, 2000 - 2500, and 3400 - 3600~cm$^{-1}$ ranges. The in-plane and out-of-plane CH bending vibrational frequencies of CN-PAHs are shifted when comparing isomers and to their unsubstituted counterparts, making their differentiation in mixed laboratory experiments possible. The overall aromatic CH stretch fundamental (2950 - 3200~cm$^{-1}$) and first overtone (5950 - 6200~cm$^{-1}$) regions are relatively unaffected by the cyano-substitution, with changes only to the breadth and intensity of the bands. Detailed spectroscopic data on the normal mode components of each state reported herein provide the means to directly assign future laboratory spectra and to guide direct IR observations of astronomical regions with, e.g., JWST.

\end{abstract}

\section{Introduction}
From the char marks on grilled meat to Earth's atmosphere and out into the cold (and hot) depths of space, polycyclic aromatic hydrocarbons (PAHs) are ubiquitous. PAHs play a fundamental role in many facets of chemistry, and this extends into chemical process that occur in space. They clump together to form the core of dust grains and ices in the interstellar medium (ISM), and act as surfaces for the catalysis of molecular reactions. There is also strong evidence that PAHs are the main carriers of the aromatic infrared bands (AIBs) detected toward most astronomical sources \cite{tielens_interstellar_2008}. The major features at 3.3, 6.2, 7.7, 11.2, 12.7, and 16.4 \textmu m are mainly attributed to fundamental CH and CC stretching and bending vibrations of PAHs \cite{tielens_interstellar_2008,chown_pdrs4all_2024,peeters_rich_2002}.

The spectroscopy of PAHs has been explored in-depth using both experimental and computational methods for almost three decades now \cite{duley_infrared_1981,allamandola_polycyclic_1985,allamandola_interstellar_1989,maltseva_high-resolution_2015,mackie_anharmonic_2015,mackie_anharmonic_2016,maltseva_high-resolution_2016,maltseva_high-resolution_2018,lemmens_anharmonicity_2019, rap_low-temperature_2022,rap_ionic_2024}. Various matrix-isolation and gas-phase experiments have been performed on PAHs of different shapes and sizes with commensurate computations to help guide and interpret the experimental data. In recent years, anharmonic vibrational methods have become computationally accessible for small PAHs. While scaled harmonic frequencies can be useful in the initial analysis of PAH spectra, the inclusion of anharmonicity in the computations is needed to accurately capture many aspects of the spectra including combination bands, overtones, and intensities, especially in the mid- to far-IR (300 - 2000~cm$^{-1}$) and aromatic CH stretches around 3000~cm$^{-1}$ \cite{esposito_anharmonic_2023,esposito_anharmonicity_2023,mackie_anharmonic_2015,mackie_anharmonic_2016,mackie_anharmonic_2018,mackie_fully_2018,mackie_anharmonicity_2022,banhatti_formation_2022,pirali_high-resolution_2009,mulas_anharmonic_2018,esposito_infrared_2024}. Whereas anharmonic computations provide a more truthful assessment of the mid- to far-IR, anharmonicity is required for computation of vibrational states in the near-IR (frequencies above 2900~cm$^{-1}$) \cite{allamandola_pah_2021,esposito_assigning_2024} because overtones and combination bands are the only vibrational states that populate this region. A recent study shows that anharmonic computations can even reproduce the band shape of the experimental absorption spectra of benzene and naphthalene in the CH stretch first overtone region with good frequency agreement \cite{esposito_assigning_2024}.

Recently, the detection of cyanobenzene and cyanonaphthalene in the molecular cloud TMC-1 via radioastronomy confirms the existence of cyano-substituted PAHs (CN-PAHs) \cite{mcguire_detection_2018,mcguire_detection_2021}, molecules where a CN group replaces one of the peripheral hydrogens, in astronomical sources. This, in combination with the launch of JWST, has led to a surge in interest in the spectroscopy of CN-PAHs \cite{li_infrared_2024,li_infrared_2024-1,esposito_cn_2024,boersma_jwst_2023,agundez_aromatic_2023,vats_rotational_2022,mcguire_early_2020,cooke_benzonitrile_2020}, particularly for the CN stretch region ($\sim$2300~cm$^{-1}$). A few laboratory studies have investigated the absorption spectrum of cyanobenzene, cyanonaphthalene, and cyanoanthracene in Ar- and H$_2$O-matrices \cite{bernstein_infrared_1997,kwon_vibrational_2003} as well as in the gas-phase \cite{rajasekhar_spectroscopic_2022}, but correlation to interstellar environments is left wanting. Recent work reports the computational anharmonic absorption spectra for a handful of small CN-PAHs in the CN stretch region: cyano-substituted benzene, naphthalene, anthracene, pyrene, and phenanthrene \cite{esposito_cn_2024}. These computations reveal a narrow, strong band in the 2270 - 2310~cm$^{-1}$ (4.405 - 4.329 \textmu m) range that does not change with size or shape, and strong anharmonic coupling that leads to a complex spectrum. This band is completely unique to PAHs with A CN substituent. In the present work, the remainder of the absorption spectra (300 - 2000~cm$^{-1}$ and 2500 - 6200~cm$^{-1}$) of these CN-PAHs are computed to provide a guide for high-resolution experiments on these molecules. Additionally, the results generate a much more complete understanding of how astronomical CN-PAH IR emission may look within the high-resolution spectra captured by JWST. 

\section{Computational Methods}

The optimized geometry, harmonic frequencies and normal modes are computed with the density functional B3LYP \cite{becke_densityfunctional_1993} in conjunction with the N07D basis set \cite{barone_development_2008} using Gaussian 16 \cite{m_j_frisch_g_w_trucks_h_b_schlegel_g_e_scuseria_m_a_robb_j_r_cheeseman_g_scalmani_v_barone_g_a_petersson_h_nakatsuji_et_al_gaussian_2016}, a well-benchmarked approach \cite{maltseva_high-resolution_2015}. The geometry optimizations use a custom integration grid consisting of 200 radial shells and 974 angular points per shell \cite{mackie_anharmonic_2015}. The CN-PAHs considered herein are the singly-substituted: cyanobenzene (C$_6$H$_5$CN), cyanonaphthalene (C$_{10}$H$_7$CN), cyanoanthracene (C$_{14}$H$_{9}$CN), cyanophenanthrene (C$_{14}$H$_{9}$CN), and cyanopyrene (C$_{16}$H$_{9}$CN) in all of their symmetry-unique isomers shown in Figure~S1. The N07D basis set is based on the double-$\zeta$ 6-31G(d) basis with additional diffuse and polarization functions added that have been shown to better treat anharmonicity in large, aromatic systems such as PAHs \cite{barone_fully_2014}. 

Following optimization, the quadratic, cubic, and quartic normal coordinate force constants (quartic force field; QFF) are computed using B3LYP/N07D. A QFF is a fourth order Taylor series expansion of the potential energy surface around the equilibrium geometry with the following formula:
\begin{equation}
  \begin{split}
     V = & \frac{1}{2}\sum_{i,j}^{3N} {\biggl (\frac{\partial^2 V}{\partial X_i\partial X_j}\biggl)X_iX_j} \\
     + & \frac{1}{6}\sum_{i,j,k}^{3N} {\biggl (\frac{\partial^3 V}{\partial X_i\partial X_j\partial X_k}\biggl)X_iX_jX_k} \\
     + & \frac{1}{24}\sum_{i,j,k,l}^{3N} {\biggl (\frac{\partial^4 V}{\partial X_i\partial X_j\partial X_k\partial X_l}\biggl)X_iX_jX_kX_l}, \\
  \end{split}
\end{equation}
The force constants are computed through small displacements of atoms along predefined normal mode coordinates. The QFF is then transformed into Cartesian coordinates via a linear transformation \cite{mackie_linear_2015}.

Second order vibrational perturbation theory (VPT2) \cite{fortenberry_chapter_2019,franke_how_2021,watson_vibrational_1977,mills_molecular_1972,bowman_vibrational_2022} is used to compute the anharmonic vibrational frequencies of each CN-PAH isomer with a modified version of the \textsc{Spectro} \cite{gaw_spectro_1991} software. \textsc{Spectro} has the advantage of implementing polyad resonance matrices in the VPT2 treatment \cite{martin_accurate_1997,martin_anharmonic_1995}. When vibrational states of the same symmetry are close in energy, they interact via resonance coupling (e.g., Fermi, Darling-Dennison). When the density of vibrational states is high, each state can participate in multiple resonances at once, termed ``resonance chaining.'' Using the resonance polyads within \textsc{Spectro} allows for simultaneous treatment of these resonances as well as accounting for the redistribution of the vibrational oscillator strength, phenomena that are difficult to properly treat with standard VPT2 implementations. In this study, states with a frequency separation of 200~cm$^{-1}$ or less are included in the polyads\cite{mackie_anharmonic_2015}. Vibrational modes with frequencies below 300~cm$^{-1}$ are excluded from the VPT2 treatment due to known issues in accurately describing their potential energy surfaces.

The resulting anharmonic vibrational frequencies are convolved with a Lorentzian profile having a full-width at half-maximum (FWHM) of 1 and/or 15~cm$^{-1}$ in order to create a continuous spectrum for each of the molecules considered. 

\section{Results and Discussion}

In general, cyano-substitution leads to: 1) lower symmetry, and therefore more IR active modes, 2) a change in the vibrational normal modes, 3) a shift in the charge balance leading to different and larger intensities, 4) new, distinct features associated with the CN group, and 5) different anharmonic coupling. The implications of these changes will be explored in the following discussion. 

\begin{figure}
    \centering
    \includegraphics{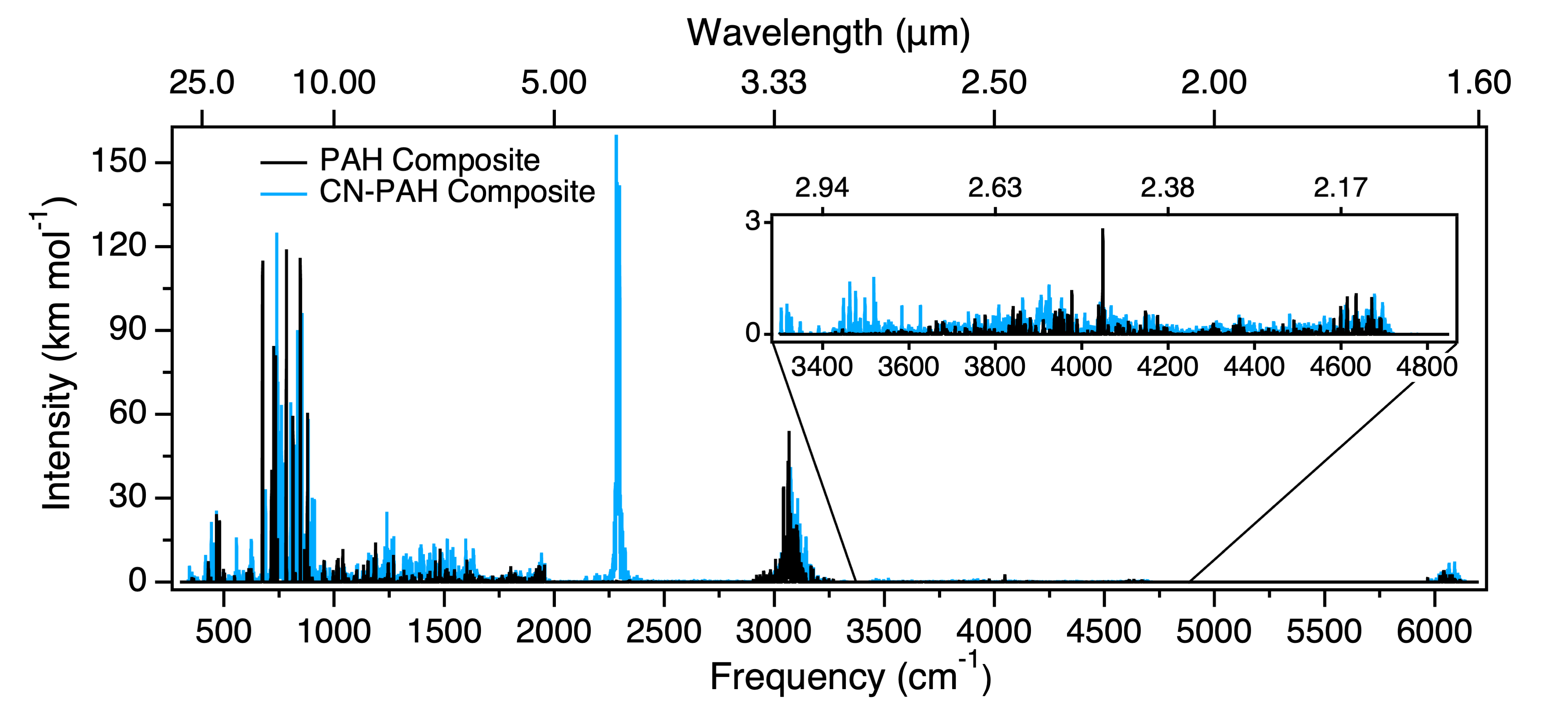}
    \caption{Composite computed anharmonic spectra of cyano-substituted PAHs (blue) and standard PAHs (black) from 300 - 6200~cm$^{-1}$. The inset displays an expanded view of the region from 3350 - 4850~cm$^{-1}$.}
    \label{fig:total_composite}
\end{figure}

Figure~\ref{fig:total_composite} presents the composite anharmonic absorption spectrum from 300 - 6200~cm$^{-1}$ of the PAHs and CN-PAHs studied here. Because the data presented here are applicable both to laboratory experiments and observations, these composite spectra are produced by summing the individual spectrum of each isomer of each molecule instead of averaging them. The detailed data for each individual isomer is available in the Supplementary Material. The most striking difference between the PAH and CN-PAH spectra is the presence of a intense, narrow CN-PAH band centered around 2300~cm$^{-1}$. This feature originates from the CN stretching fundamentals and other modes coupled to the CN stretch. Due to its relevance to astrophysics, this feature is described in greater detail in a separate publication \cite{esposito_cn_2024}. Other regions of the spectra show similarities and differences, in particular in the 700 - 950, 1100 - 1300, 2950 - 3200, 3400 - 4800, and 5950 - 6200 ~cm$^{-1}$ ranges. The current discussion will focus on the analysis of these portions of the spectra.

\subsection{Out-of-Plane CH Bend Region (700 - 950~cm$^{-1}$)}

\begin{figure}
    \centering
    \includegraphics{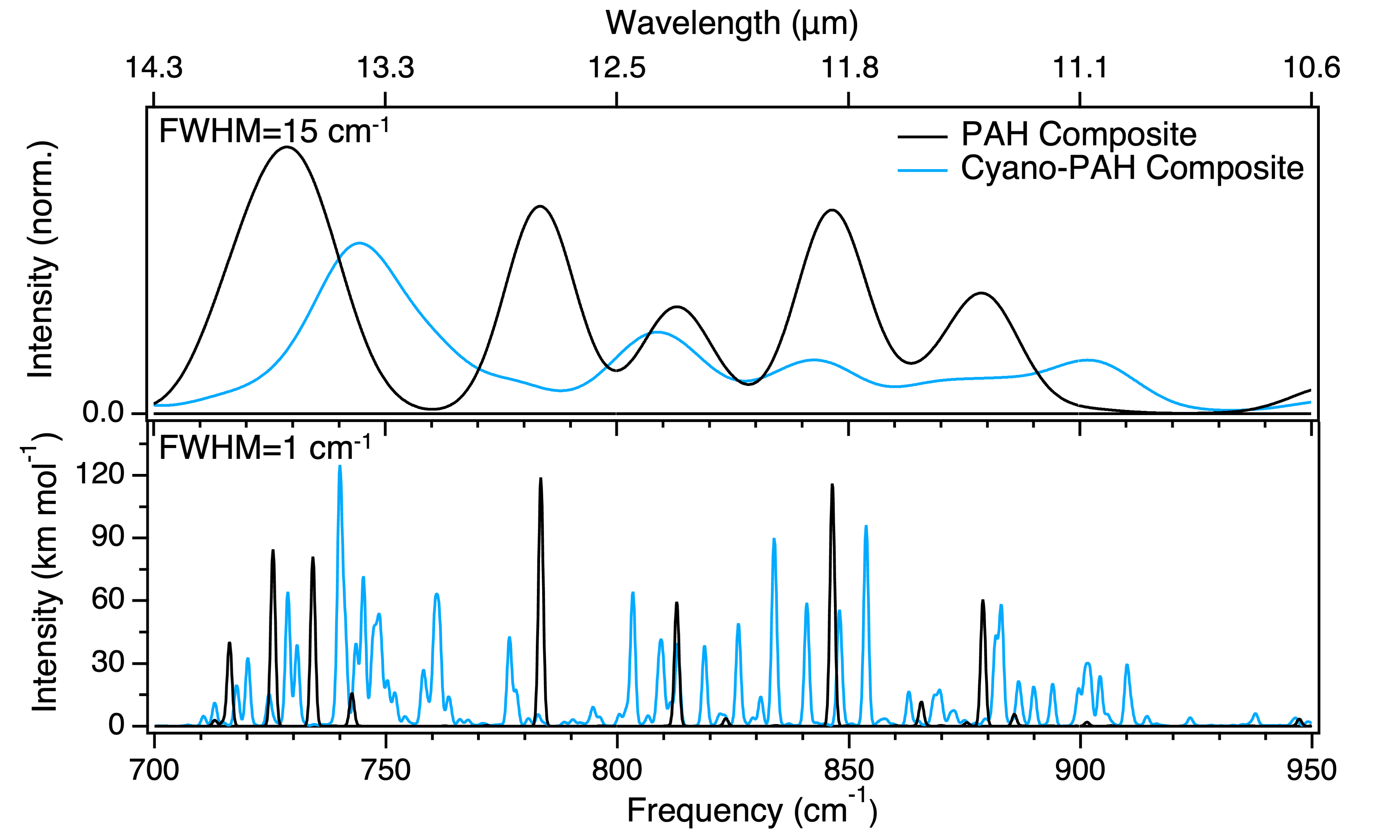}
    \caption{Composite computed anharmonic spectra of cyano-substituted PAHs (blue) and standard PAHs (black) from 700 - 950~cm$^{-1}$, the out-of-plane CH bending region, broadened using a Lorentzian line shape with a FWHM of 15~cm$^{-1}$ (top) and 1~cm$^{-1}$ (bottom).}
    \label{fig:oop_high_low_res}
\end{figure}

Figure~\ref{fig:oop_high_low_res} focuses on the 700 - 950~cm$^{-1}$ region of the spectrum. In standard PAHs, this frequency range is dominated by out-of-plane (OOP) CH bending motions; a behavior also shown here in the CN-PAHs. When a narrow FWHM of 1~cm$^{-1}$ is applied (the bottom panel of Figure~\ref{fig:oop_high_low_res}), a higher density of IR active vibrational transitions is observed in the CN-PAH spectrum compared to the PAH spectrum. The higher density of transitions is partly attributed to the presence of a greater number of isomers resulting from the addition of the CN substituent and partly to the lower symmetry created by the addition of the CN group. 


\begin{figure}
    \centering
    \includegraphics{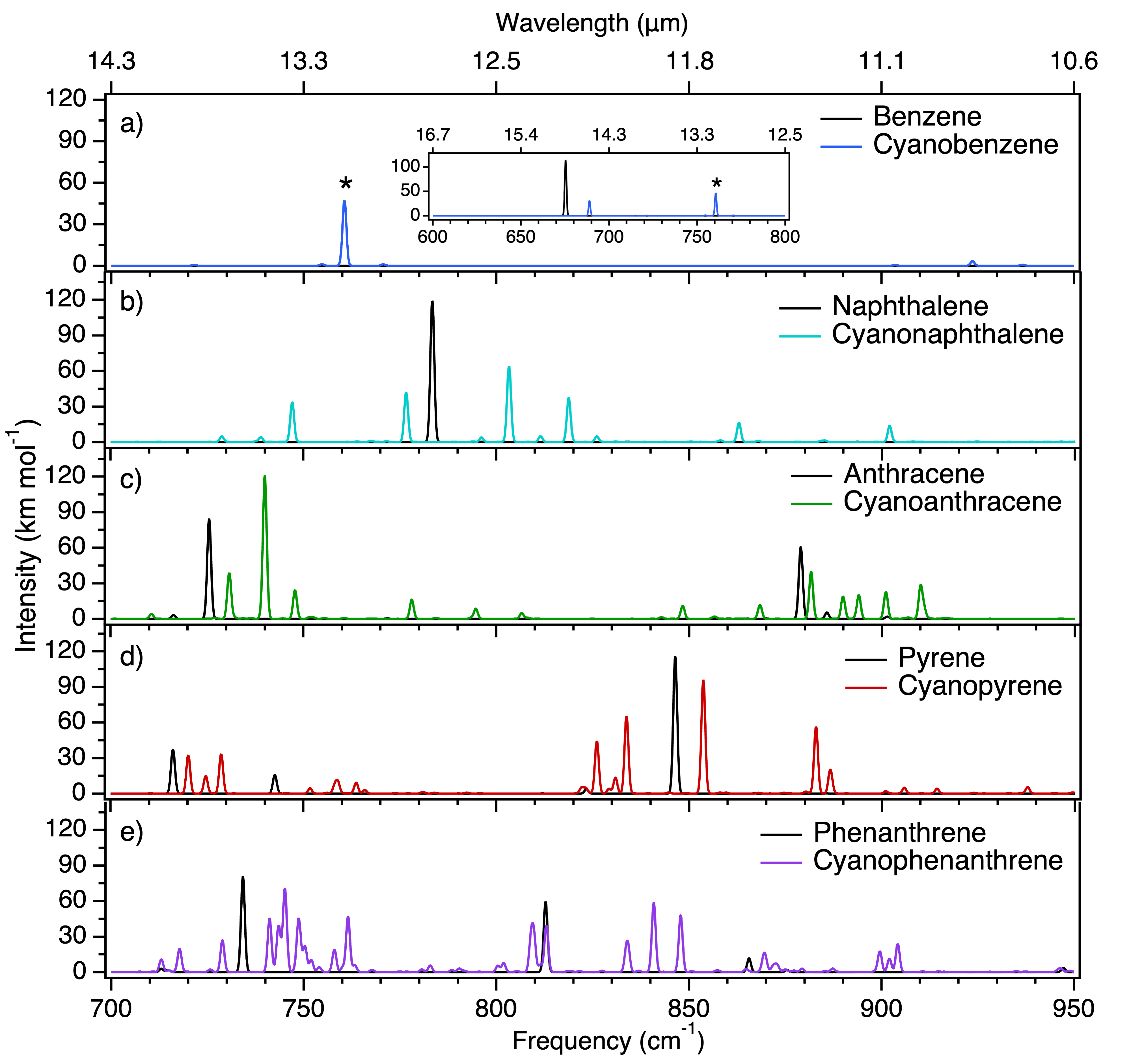}
    \caption{Composite anharmonic absorption spectra from 700 - 950~cm$^{-1}$, the out-of-plane CH bending region, for all isomers of each molecule including the (a) one isomer of cyanobenzene, (b) two isomers of cyanonaphthalene, (c) three isomers of cyanoanthracene, (d) three isomers of cyanopyrene, and (e) five isomers of cyanophenanthrene.}
    \label{fig:cyano_individual_molecs}
\end{figure}

Figure~\ref{fig:cyano_individual_molecs} presents the 700 - 950~cm$^{-1}$ absorption spectra of the sum of all isomers for each of the individual CN-PAHs broadened with a FWHM of 1~cm$^{-1}$. These spectra provide insight into the detailed structure of each molecule as well as provide guidance and reference data for future laboratory studies and observations. The intensities in this region are large, driven by the OOP CH bending motions that distort the molecules out of the molecular plane causing a large change in the dipole moment. The spectrum of benzene and the single isomer of cyanobenzene (panel (a)) is sparse. The panel (a) inset shows slightly shifted frequencies between 600 - 800~cm$^{-1}$.  Because two of the three peaks in these molecules are found here, the same feature is marked with an asterisk for clarity. The single peak in the benzene spectrum occurs at 675.4~cm$^{-1}$ and is the symmetric OOP CH bend. Two peaks are visible in the cyanobenzene spectrum, one of which is the asymmetric OOP CH bending motion at 688.9~cm$^{-1}$, and the other the symmetric OOP CH bend at 760.6~cm$^{-1}$. The two OOP bends in cyanobenzene are approximately 40\% and 25\% as intense (47 and 31~km mol$^{-1}$), respectively, as the same motion in benzene (115~km mol$^{-1}$).

The remainder of the panels depict the spectra of cyanonaphthalene (two isomers, Panel (b)), cyanoanthracene (three isomers, Panel (c)), cyanopyrene (three isomers, Panel (d)), and cyanophenanthrene (five isomers, Panel (e)) in the OOP CH bending region. In naphthalene, there is a single strong feature at 783.5~cm$^{-1}$ belonging to the OOP CH bend. Anthracene has two intense peaks, pyrene has three, and phenanthrene has two. The spectra of the standard PAHs are sparse partly due to having high symmetry, including a few OOP CH bending fundamentals and not much other activity. Adding the cyano group increases the level of complexity in the OOP CH bend region for each system. For example, the cyanopyrene spectrum has six features with intensities of greater than 30~km mol$^{-1}$. The most intense peak for pyrene at 846.4~cm$^{-1}$ is the symmetric OOP CH bend. Two of the intense peaks in the cyanopyrene spectrum belong to 1-cyanopyrene. They both are OOP CH bends, but the addition of the CN group disrupts the vibrational normal modes and affects the spectrum. The peak at 882.9~cm$^{-1}$ (56~km mol$^{-1}$) arises from bending of the CH bonds closest to the CN group, while the weaker peak at 826.1~cm$^{-1}$ (44~km mol$^{-1}$) involves bending of the CH bonds furthest from the CN group. Consequently, the addition of the CN functional group to the apex of pyrene in the $C_{2v}$ CN-PAH isomer splits the symmetric OOP CH bend into two weaker features at higher and lower frequency.

A single peak occurs at 734.2~cm$^{-1}$ in the standard phenanthrene spectrum that belongs to the asymmetric OOP CH bend where the quartet of CH bonds are bending in unison and the duo CHs are opposing. The cyanoanthracene spectrum includes a cluster of IR active vibrational states around 740 - 760~cm$^{-1}$ that does not exist in the phenanthrene spectrum. Within this range, there are 44 vibrational states between the five isomers, 14 of which have an intrinsic intensity of greater than 1~km mol$^{-1}$. This leads to the build-up of a relatively broad and structured feature as compared to the sharp lines throughout the other portions of this spectrum and the spectra of the other molecules. As has been shown in previous studies \cite{esposito_anharmonicity_2023,esposito_anharmonic_2023,esposito_infrared_2024,mackie_anharmonic_2015,mackie_fully_2018,mulas_anharmonic_2018,esposito_assigning_2024}, the anharmonic computations shown here can be used to directly identify the spectral carrier in laboratory studies as well as provide a detailed assignment of the individual features. In this frequency range in particular, the computations have enough precision to allow for isomeric identification based on frequency shifts between isomers within each family because the sparsity of peaks allows for requisite separation. Additionally, the ability to identify frequency shifts in the CN stretch fundamental region (2000 - 2500~cm$^{-1}$) between the same isomers included in this study was previously demonstrated \cite{esposito_cn_2024}.

Analysis at a FWHM of 15~cm$^{-1}$ (top panel, Figure~\ref{fig:oop_high_low_res}) shows the composite spectrum of the PAHs and CN-PAHs from 700 - 950~cm$^{-1}$ at a width that might be expected in an astronomical spectrum. When broadened, the spectra of the two groups of molecules share more similarity, but some differences persist. There are five distinct, fairly narrow peaks in the PAH spectrum, whereas in the CN-PAH spectrum there are four broader, flatter peaks. This is a result of the lone, strong peaks in the PAH spectrum, and the more congested, broad range of peaks in the CN-PAH spectrum. The most intense peak in the PAH spectrum is centered at $\sim$730~cm$^{-1}$ while the strongest CN-PAH feature is at $\sim$740~cm$^{-1}$. Overall, the CN-PAH spectrum more resembles a plateau compared to the spiky mountain range-like structure of the pure PAH spectra. However, this is only a small PAH sample. The blending of lines from potentially hundreds or even thousands of large PAHs is currently thought to produce the strong astronomical infrared bands detected between 11 - 15 \textmu m. Based on the results from the current study, CN-PAHs could possibly contribute to the intensity of the 11 - 15 \textmu m astronomical PAH bands and particularly to the underlying 900-660~cm$^{-1}$ plateau. Due to the blending and overlapping nature of these bands, differentiation between PAHs and CN-PAHs in this region of the spectrum would be impossible, complicating declaration of the true spectral carrier of these bands as pure PAHs, functionalized PAHs, or a combination of both. Although, the verified presence of CN-PAHs in space by radio observations make a strong case for CN-PAHs possibly contributing to the 11 - 15 \textmu m emission in other types of astronomical objects. 



\subsection{In-Plane CH Bend Region (1100 - 1300~cm$^{-1}$))}

\begin{figure}
    \centering
    \includegraphics{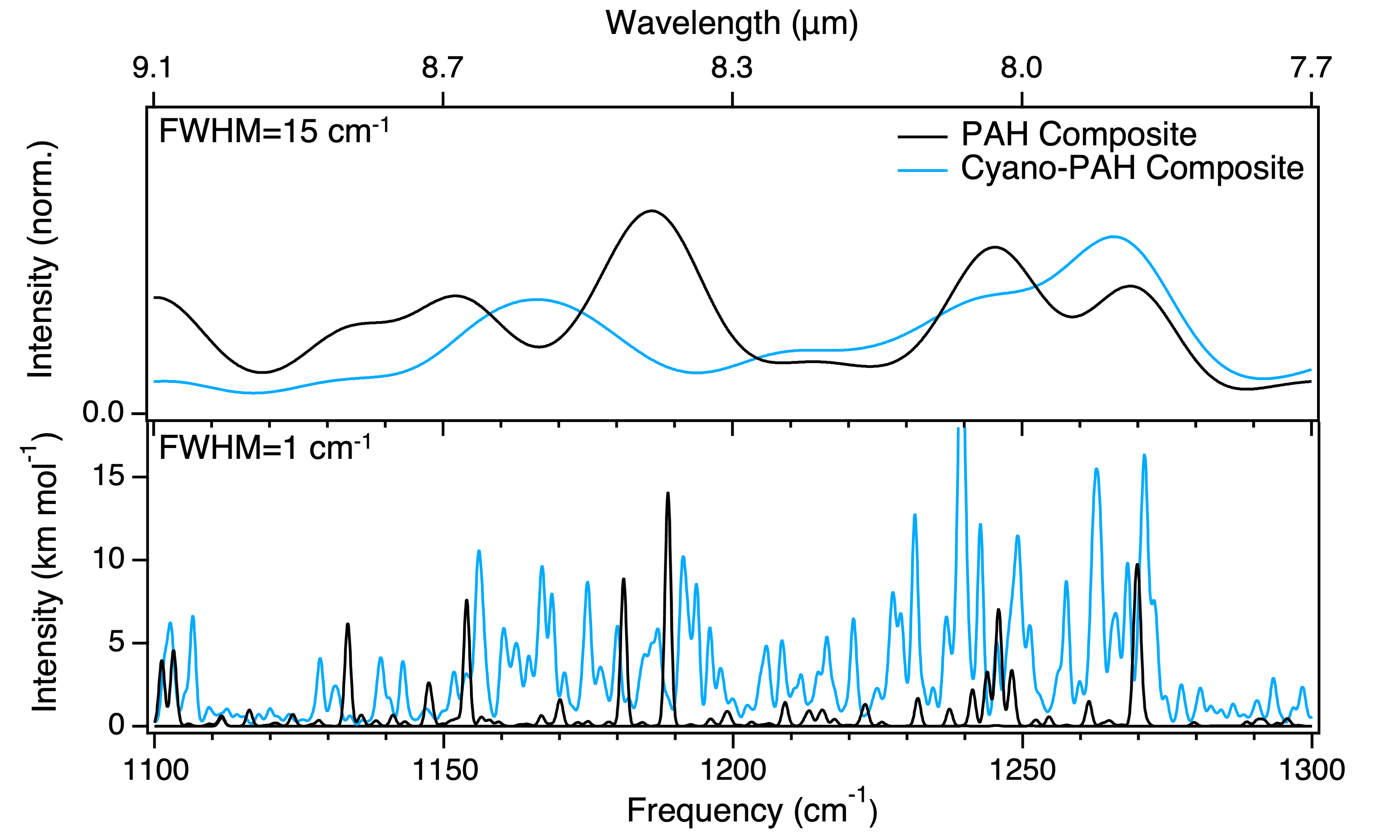}
    \caption{Composite computed anharmonic spectra of cyano-substituted PAHs (blue) and standard PAHs (black) from 1100 - 1300~cm$^{-1}$, the in-plane CH bending region, broadened using a Lorentzian line shape with a FWHM of 15~cm$^{-1}$ (top) and 1~cm$^{-1}$ (bottom).}
    \label{fig:inplane_high_low_res}
\end{figure}

Figure~\ref{fig:inplane_high_low_res} depicts the composite CN-PAH and standard PAH spectra from 1100 - 1300~cm$^{-1}$ broadened with a FWHM of 15~cm$^{-1}$ (top) and 1~cm$^{-1}$ (bottom). The higher-resolution spectra show a more crowded, dense, and intense spectrum for the CN-PAHs than the standard PAHs. Whereas the standard PAH spectrum has a few sharp peaks with relatively little intensity in between, the CN-PAH spectrum has a jagged peak structure with almost constant intensity from 1150~cm$^{-1}$ to 1270~cm$^{-1}$. While this is an intriguing difference on the scale of a few wavenumbers, it may not have an impact when any sort of broadening occurs. For example, when the spectra are broadened to a FWHM of 15~cm$^{-1}$ and normalized to their tallest peak, no unique, identifiable structure is retained in the CN-PAH spectrum compared to the PAH spectrum. Both sets of molecules overlap over the entire range leading to a similar conclusion as in the OOP CH stretch region: CN-PAHs may contribute to the strong aromatic infrared bands from 7 - 9 \textmu m, but there is not a unique feature in this spectral range that will allow for the unambiguous determination of the presence of CN-PAHs.

\begin{figure}
    \centering
    \includegraphics{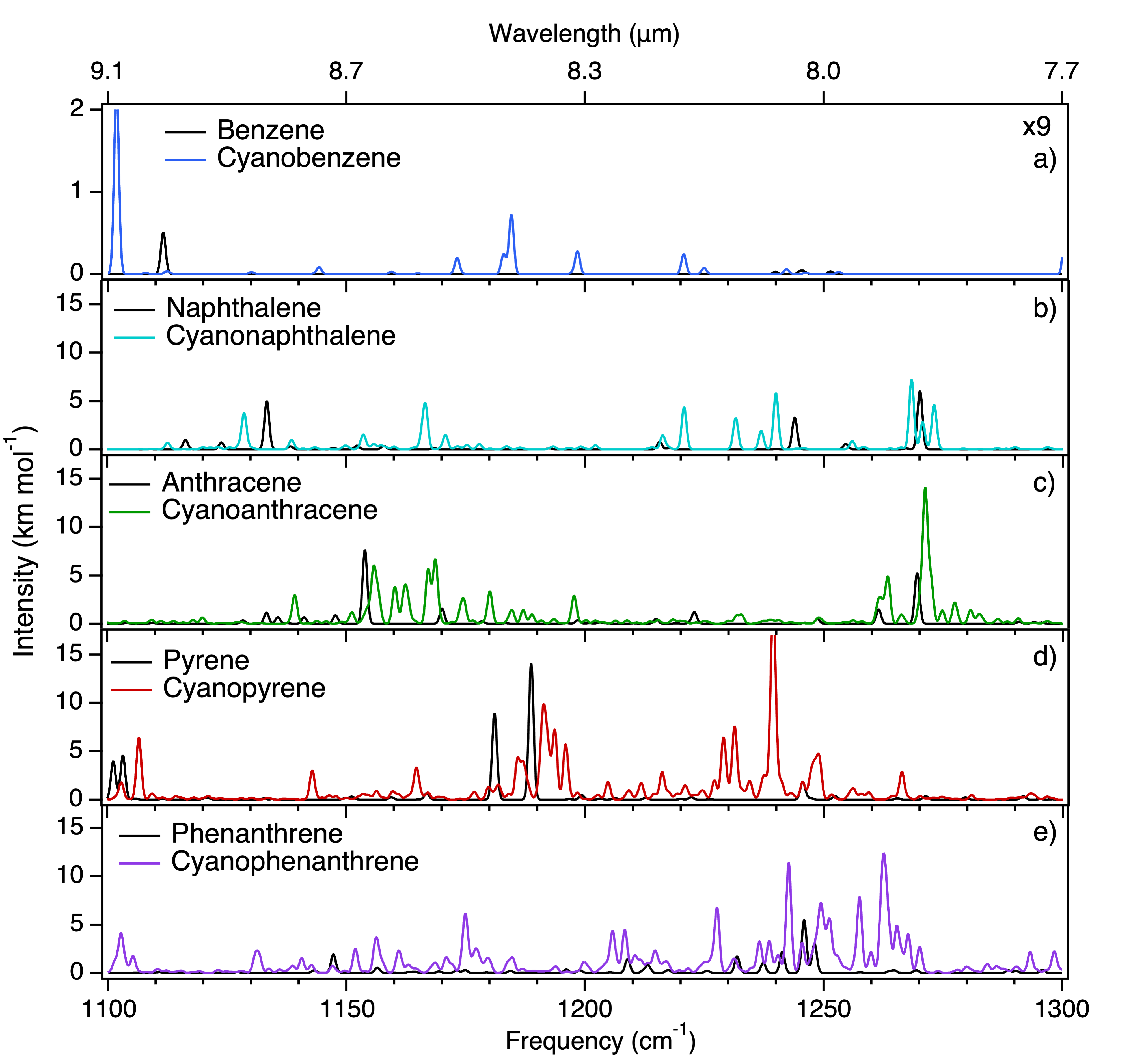}
    \caption{Composite anharmonic absorption spectra from 1100 - 1300~cm$^{-1}$, the in-plane CH bending region, for all isomers of each molecule, including the (a) one isomer of cyanobenzene, (b) two isomers of cyanonaphthalene, (c) three isomers of cyanoanthracene, (d) three isomers of cyanopyrene, and (e) five isomers of cyanophenanthrene.}
    \label{fig:in_plane_5stack}
\end{figure}

Figure~\ref{fig:in_plane_5stack} presents the composite, anharmonic absorption spectrum for each group of molecules in the 1100 - 1300~cm$^{-1}$ frequency range. This region is often attributed to the in-plane (IP) CH bending vibrations of PAHs. The peaks in this region have an intensity approximately an order of magnitude lower than in the OOP CH bend region. While the dipole shift from the OOP motions causes large intensities in the 700 - 950~cm$^{-1}$ region, the IP motions here lead to only a relatively small change in the dipole moment and, consequently, weaker intensities. The unsubstituted PAH spectra for all molecules are equally as sparse as in the 700 - 950~cm$^{-1}$ range, which is to be expected when considering there is potentially one strong IP and one strong OOP CH bend for each molecule. In contrast, the spectra of the cyano-substituted PAHs are complex and show major differences both when compared to the standard PAHs within the family as well as across the range of PAH structures studied here. The spectra of cyanopyrene and cyanophenanthrene exhibit the most complexity and so are discussed in more detail below.

When comparing spectra of related molecules, the main goal is to identify unique, differentiating areas of the spectrum to act as a sort of fingerprint. This is most notably the case in the cyanopyrene spectrum, although, even pyrene has some tricks up its sleeve. Pyrene has two strong peaks, namely at 1181.1 and 1188.7~cm$^{-1}$. The peak at 1188.7~cm$^{-1}$ is the expected strong IP CH bend, $\nu_{30}$. What is unexpected is the second strong peak in this region. Interestingly, the peak at 1181.1~cm$^{-1}$ is a relatively intense combination band consisting of two, opposing asymmetric OOP CH bends, $\nu_{55}$ and $\nu_{60}$. Neither of these modes have much fundamental intensity, but when included in the resonance polyad treatment, the combination band steals intensity from the nearby, strong IP CH bend. This is another example of the power of the resonance polyad method and its necessity for computing accurate vibrational intensities. This is especially the case for molecules that have a large number of fundamental vibrational modes.

The cyanopyrene spectrum (panel (d) in  Figure~\ref{fig:in_plane_5stack}) shows peaks interspersed throughout the full range and two regions of higher density and stronger intensity. The region of most interest is the group of peaks from 1220 - 1250~cm$^{-1}$. There is only one small peak in the pyrene spectrum here, making it a spectral region that is completely unique to the cyanopyrene isomers when comparing the functionalized and non-functionalized PAHs. There are 92 vibrational states within the 1220 - 1250~cm$^{-1}$ range when considering all three isomers of cyanopyrene. The two peaks at 1229.0 (a mixture of $\nu_{28}$ and $\nu_{38}$+$\nu_{70}$ in 3-cyanopyrene) and 1231.3~cm$^{-1}$ (a mixture of $\nu_{28}$ and $\nu_{51}$+$\nu_{61}$ in 2-cyanopyrene) make up the majority of the intensity of the cluster from 1225 - 1235~cm$^{-1}$. The underlying baseline is comprised of 31 other weak, one- and two-quanta modes.

The peak with the strongest intensity in the entire cyanopyrene spectrum from 1100 - 1300~cm$^{-1}$ is centered at 1239.3~cm$^{-1}$ and spans from 1235 - 1245~cm$^{-1}$. The intensity leading to the tall central peak belongs to 2-cyanopyrene and is a product of strong anharmonic coupling between multiple states, as there are two overlapping strongly mixed states here. The two states are centered at 1239.1 and 1239.5~cm$^{-1}$. They are composed of the same three vibrational modes, $\nu_{46}$+$\nu_{65}$, $\nu_{44}$+$\nu_{66}$, and $\nu_{27}$, mixed in different proportions. The state at 1239.1cm$^{-1}$ has majority contributions of 35/31/12\% of $\nu_{46}$+$\nu_{65}$, $\nu_{44}$+$\nu_{66}$, and $\nu_{27}$, respectively, with the balance filled by minor contributions from a host of other modes. The state at 1239.5~cm$^{-1}$ is less mixed, with a majority contribution of 62/13/6\% for the involved states. Similar to the clustering at lower frequency, the underlying baseline and shoulders are made up of 27 other vibrational states. This level of complexity should be easily discernible in any laboratory absorption spectrum, with the results presented in this work  allowing for a full assignment.

The isomers of cyanophenanthrene (panel (e) in Figure~\ref{fig:in_plane_5stack}) have an equally complex spectrum, although phenanthrene has a few peaks that overlap with the dense range of cyanophenanthrene bands from 1230 - 1250~cm$^{-1}$. Immediately past 1250~cm$^{-1}$ is the strongest peak that fortuitously does not have any overlapping intensity from phenanthrene. From the dip in intensity at 1255~cm$^{-1}$ to the end of the jagged oscillations at 1272~cm$^{-1}$ there are 70 total vibrational states; three of which have appreciable intensity. The narrow peak at 1257.6~cm$^{-1}$ (7.7~km mol$^{-1}$) is composed of the modes $\nu_{25}$, 2$\nu_{51}$, and $\nu_{49}$+$\nu_{53}$ from 3-cyanophenanthrene, with the intensity derived from $\nu_{25}$. There are two overlapping peaks at 1262.2~cm$^{-1}$ (3-cyanophenanthrene with a majority contribution from $\nu_{48}$+$\nu_{56}$) and 1262.9~cm$^{-1}$ (1-cyanophenanthrene with a majority contribution from $\nu_{32}$+$\nu_{66}$). Although they have almost identical frequencies, their intensities are derived from different coupling partners. The peak at 1262.2~cm$^{-1}$ grabs intensity from $\nu_{25}$ (asymmetric IP CH bend) and $\nu_{23}$ (CC skeletal stretch), while the peak at 1262.9~cm$^{-1}$ derives its intensity almost exclusively from $\nu_{24}$ (asymmetric IP CH bend). 

\subsection{Aromatic CH Stretch Region (2950 - 3200~cm$^{-1}$)}

\begin{figure}
    \centering
    \includegraphics{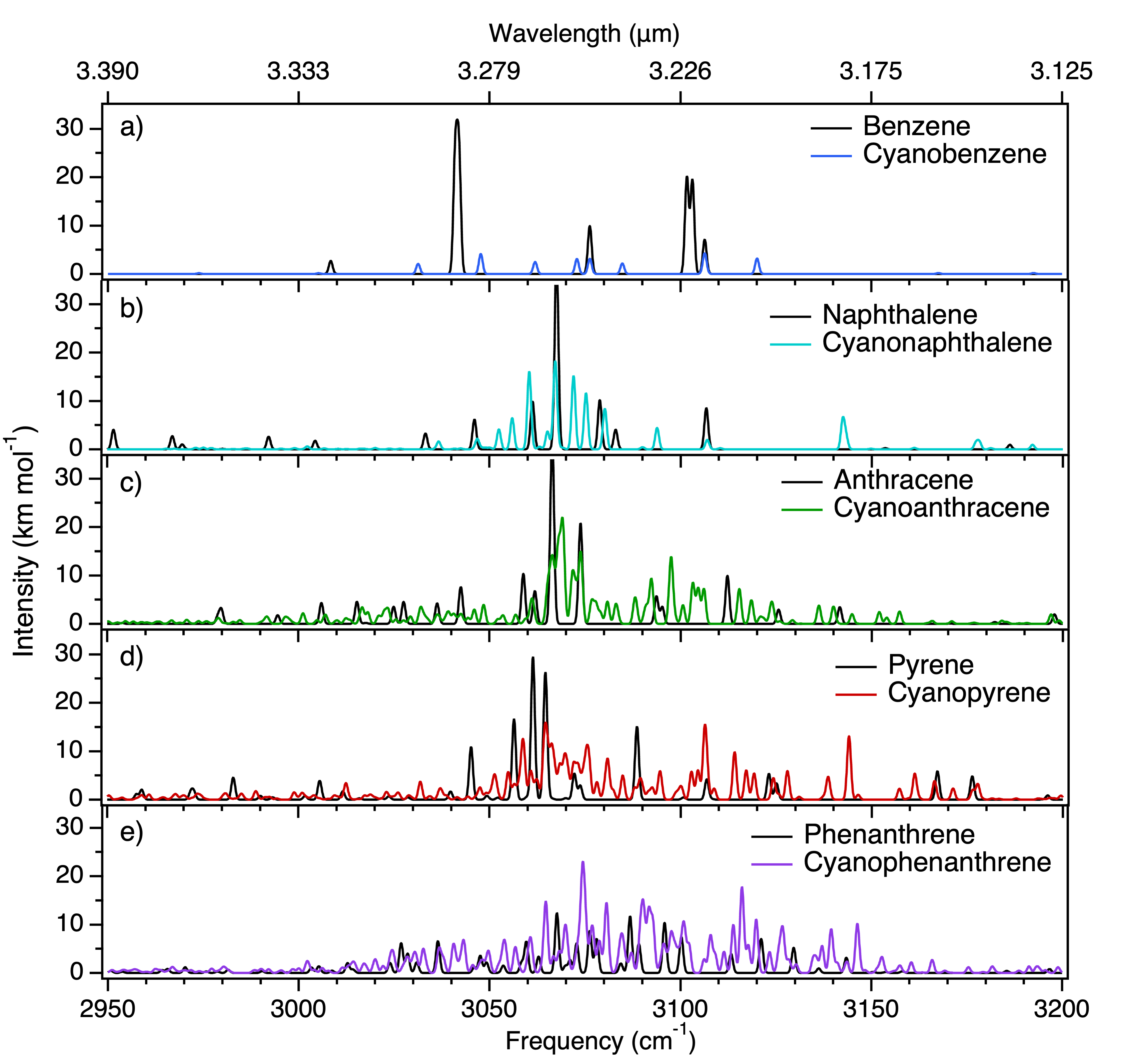}
    \caption{Composite anharmonic absorption spectra from 2950 - 3200~cm$^{-1}$ for all isomers of each molecule including the (a) one isomer of cyanobenzene, (b) two isomers of cyanonaphthalene, (c) three isomers of cyanoanthracene, (d) three isomers of cyanopyrene, and (e) five isomers of cyanophenanthrene.}
    \label{fig:arom_ch_5panel}
\end{figure}

Figure~\ref{fig:arom_ch_5panel} presents the anharmonic absorption spectrum of the composite of each family of molecules in the 2950 - 3200~cm$^{-1}$ frequency range. These peaks arise from the aromatic CH stretching motions of the hydrogens that decorate the periphery of the PAHs. Because these peaks stem from the CH stretches and do not directly involve the CN functional group, the expectation is that the general motif of the spectrum will stay the same. This hypothesis is confirmed when inspecting Figure~\ref{fig:arom_ch_5panel} even though there are some subtle differences. 

For each of the parent molecules and CN-substituted versions, the overall shape and central frequency of the peaks overlap almost exactly. Where they differ is in the total intensity and breadth of the peaks. The standard PAH spectra have mostly individual, sharp peaks. The onset of intensity begins around the same place for each PAH and CN-PAH, but the CN-PAHs have intensity that extends to higher frequency more so than the standard PAHs. 

Analysis of the individual CN-PAH spectra reveals that the additional density of peaks stems mostly from the increased number of isomers. The five isomers of cyanophenanthrene have a total integrated intensity of 502~km mol$^{-1}$ while phenanthrene has an integrated intensity of 150~km mol$^{-1}$. Further, the combination bands that include one quantum of the CN stretching mode ($\nu_{CN}$+$\nu_{x}$ = $\sim$3100~cm$^{-1}$) do not show large intensities because of small anharmonic couplings even though some combinations include the intense OOP CH bends. Indeed, if present, the aromatic CH stretches in CN-PAHs will contribute to the 3.3 \textmu m feature and potentially be a source of emission out to 3.1 \textmu m.

\subsection{CN Stretch First Overtone and Combination Band (3400 - 4800~cm$^{-1}$) and Aromatic CH Stretch First Overtone  (5950 - 6200~cm$^{-1}$) Regions}

Looking back at Figure~\ref{fig:total_composite}, the inset presents an expanded view of the 3400 - 4800~cm$^{-1}$ region. Overall the PAH and CN-PAH spectra are similar. However, from 3400 - 3600~cm$^{-1}$ there are strong bands in the CN-PAH spectrum that do not show up in the PAH spectrum, providing a region of uniqueness that is vital for differentiating between PAHs and CN-PAHs. Additionally, the CN-PAH spectrum in general has a higher density of vibrational transitions as well as larger intensities. 

\begin{figure}
    \centering
    \includegraphics{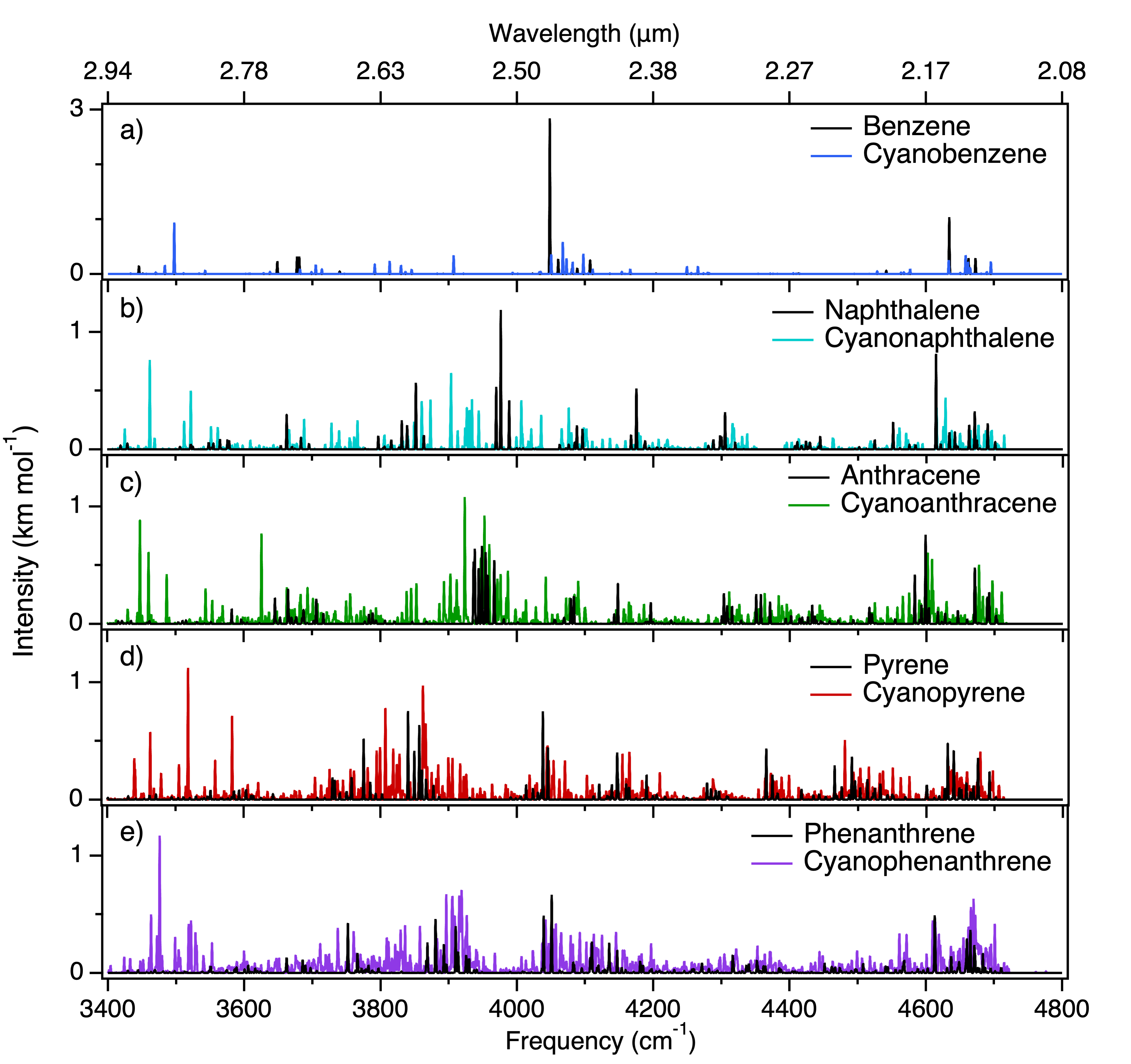}
    \caption{Composite anharmonic absorption spectra from 3400 - 4800~cm$^{-1}$, the CN stretch first overtone and combination band region, for all isomers of each molecule including the (a) one isomer of cyanobenzene, (b) two isomers of cyanonaphthalene, (c) three isomers of cyanoanthracene, (d) three isomers of cyanopyrene, and (e) five isomers of cyanophenanthrene.}
    \label{fig:cyano_individual_3400_4800}
\end{figure}

Figure~\ref{fig:cyano_individual_3400_4800} presents the composite spectrum composed of the sum of each isomer for all molecules from 3400 - 4800~cm$^{-1}$. Looking specifically at the 3400 - 3600~cm$^{-1}$ range in the cyanopyrene spectrum, three peaks jump out as having an intensity above 0.5~km mol$^{-1}$. The largest peak at 3518.45~cm$^{-1}$ is the $\nu_{28}$+$\nu_{10}$ combination band that includes a CC skeletal stretch and the CN stretch. Combination bands that include one quantum of CN stretch and one quantum of another CC skeletal stretching mode ($\nu_{10}$+$\nu_{CC}$) are often the peaks with the largest intensity in these spectra. Although, due to intensity redistribution, a different dipole moment environment created by the cyano group, and different levels of anharmonic coupling, many other modes carry higher intensity as well. This can be seen by the much larger colored peaks compared to the black ones in each spectrum. Interestingly, the number of IR active vibrational states in this region for pyrene and each individual isomer of cyanopyrene is around 95 regardless of the presence of the cyano group or substitution location. The three isomers of cyanopyrene lead to three times the number of transitions in this region, but each individually measured spectrum would have a comparable number. 

The trends described for pyrene and cyanopyrene in the 3400 - 3600~cm$^{-1}$ range hold true for all of the molecules computed here. There is slight variation in the location of the most intense peaks between species. For example, in cyanophenanthrene the intense peak at 3475.9~cm$^{-1}$ is the $\nu_{28}$+$\nu_{10}$ combination band of 2-cyanophenanthrene that includes a CC skeletal stretch and the CN stretch, analogous to the largest peak in 2-cyanopyrene at 3518.45~cm$^{-1}$. The frequency shift arises from differences in the molecular structure and substitution site. Overall, the lines in this region are very weak, and a declarative statement about their possible contribution to an astronomical IR spectrum cannot be made. However, if a fairly weak and broad astronomical feature is detected in the 2.9 - 2.7 \textmu m region, it may very well have a contribution from CN-PAHs. Additionally, high-resolution experimental studies should be able to measure these lines, and the computations provided herein can assist in the assignment of the spectra.

The remainder of the spectrum consists mostly of overlapping structure with similar intensities and patterns. The highest possible frequency of a combination band that includes a single quantum of CN stretch is $\sim$3950~cm$^{-1}$. As such, only minor variation in the spectra above this frequency is expected and, subsequently, predicted. The variation in the CN-PAH spectra can be ascribed to the overall higher number of vibrational states mentioned previously as well as frequency shifts due to a change in the vibrational normal modes. Moreover, there is an increase in the intensity of states that originate in PAHs and are retained upon cyanation due to the presence of the heavy, highly polar cyano group. Not shown here are the CN stretch plus aromatic CH stretch combination bands. These states fall in the 5300 - 5450~cm$^{-1}$ range but have extremely small intensities on the order of 0.001~km mol$^{-1}$ or less. 

\begin{figure}
    \centering
    \includegraphics{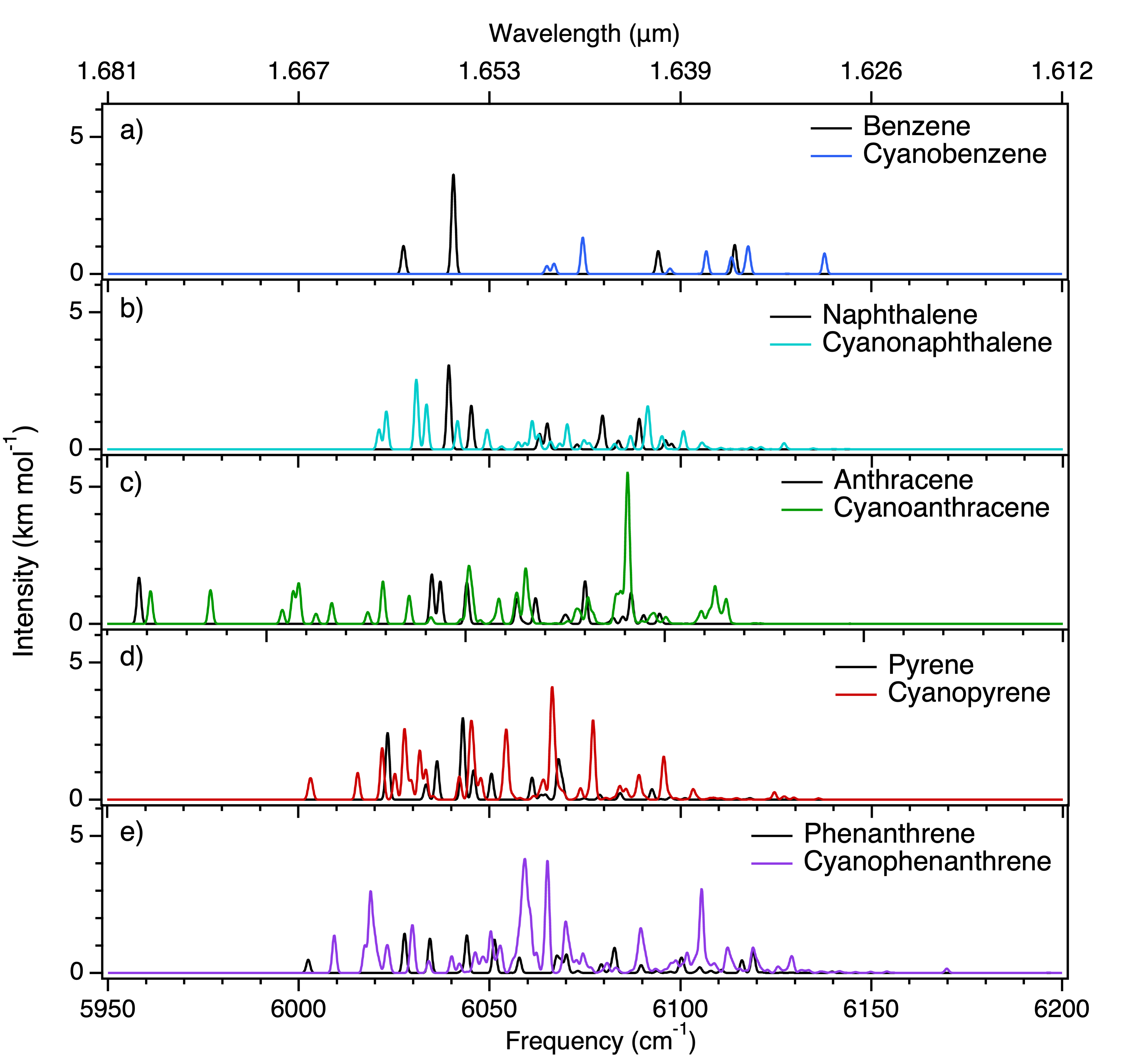}
    \caption{Composite anharmonic absorption spectra from 5950 - 6200~cm$^{-1}$, the aromatic CH stretch first overtone region,  for all isomers of each molecule including the (a) one isomer of cyanobenzene, (b) two isomers of cyanonaphthalene, (c) three isomers of cyanoanthracene, (d) three isomers of cyanopyrene, and (e) five isomers of cyanophenanthrene.}
    \label{fig:cyano_individual_5950_6200}
\end{figure}

Figure~\ref{fig:cyano_individual_5950_6200} presents the anharmonic absorption spectrum of the PAHs and CN-PAHs from 5950 - 6200~cm$^{-1}$: the aromatic CH stretch first overtones. Similarly to the CH stretch fundamentals, the overall breadth and number of features is relatively constant between the PAHs and CN-PAHs. The major difference, again, is the presence of multiple isomers for the CN-PAHs increasing the number of IR active transitions. Additionally, as has been the motif throughout these spectra, the shift in charge balance imposed by the cyano group causes an intensity increase for many of the states.

The intensity spread throughout this region comes from both aromatic CH stretch first overtones as well as 2-quanta combination bands that include different aromatic CH stretch fundamentals. For example, the intense peak in the cyanoanthracene spectrum at 6090.7~cm$^{-1}$ mainly belongs to 2$\nu_{3}$, an overtone of an asymmetric CH stretch, of 2-cyanoanthracene, while the broad peak centered at $\sim$6049~cm$^{-1}$ in cyanophenanthrene is many overlapping combination bands and overtones with relatively weak intensity from each of the five isomers. In all, though, the similar frequency range and number of IR active states between the PAHs and CN-PAHs in this region confirms the expectation that the cyano substitution does not drastically alter the vibrational structure of CH stretches in PAHs. The search for CH stretch overtone features is now possible with NIRSpec aboard JWST. Due to the more intense transitions and larger number of isomers, PAH signatures detected in the 1.6 \textmu m region \cite{geballe_detection_1994} could potentially stem from CN-PAHs in addition to PAHs.

\subsection{Conclusions}
Cyano-substituted PAHs have a rich and complex absorption spectrum across the entire 300-6200 cm-1 (3.33-1.61 \textmu m region. This paper explores the wealth of data on these molecules for the first time. Anharmonic quantum chemical computations reveal varying peak structure in different regions of the infrared dependent on molecular shape, size, and substitution site. When compared to their unsubstituted parent molecules, spectral differences are widespread in high-resolution (FWHM$\sim$1~cm$^{-1}$), most prominently in the 2200 - 2400~cm$^{-1}$ region \cite{esposito_cn_2024}, but get washed out when a broader line width (15~cm$^{-1}$) is applied. In these broad spectra, there are little-to-no unique identifying features for CN-PAHs as compared to standard PAHs, making the contribution of CN-PAHs to astronomical spectra difficult to assess definitively. Even so, and based on the results of this study, CN-PAHs may contribute uniquely beyond their pure PAH counterparts to the 11 - 15, 7 - 9, and 3.3 - 3.4 \textmu m astronomical bands based on the profiles of their IP and OOP CH bending vibrations and aromatic CH stretches. CN-PAHs may also contribute to emission detected immediately to shorter wavelengths from the 3.3 \textmu m feature.

Based on the previous success in assigning experimental spectra of PAHs using these methods \cite{esposito_anharmonic_2023,esposito_anharmonicity_2023,esposito_infrared_2024,mackie_anharmonic_2015,mackie_anharmonic_2018,mackie_anharmonicity_2022}, there is confidence that similar levels of success can be achieved for CN-PAHs. However, laboratory spectra are required to benchmark the accuracy of the computational methods used here. Very few experimental studies have been completed on these cyano-PAHs, and these data will hopefully spur new laboratory studies. Finally, to provide a better understanding of the spectroscopy and reaction chemistry that CN-PAHs take part in throughout space, these methods need to be extended to the IR emission from CN-PAHs. Li et al. have begun this work \cite{li_infrared_2024-1}, and more is needed.

\begin{acknowledgement}
V.J.E. acknowledges an appointment to the NASA Postdoctoral Program at NASA Ames Research Center, administered by the Oak Ridge Associated Universities through a contract with NASA. R.C.F. acknowledges funding from NASA grant NNH22ZHA004C and the University of Mississippi's College of Liberal Arts. C.B. is grateful for an appointment at NASA Ames Research Center through the San Jos\'{e} State University Research Foundation (80NSSC22M0107). V.J.E, R.C.F., C.B., and L.J.A. acknowledge support from the Internal Scientist Funding Model (ISFM) Laboratory Astrophysics Directed Work Package at NASA Ames. Computer time from the Pleiades and Aiken clusters of the NASA Advanced Supercomputer (NAS) is gratefully acknowledged.

\end{acknowledgement}

\begin{suppinfo}

Molecular structures for each isomer, detailed spectroscopic assignment of each vibrational state for all isomers of each molecule, and the convolved spectrum for all isomers.

\end{suppinfo}

\bibliography{references}

\end{document}